# Evaluation of High-speed Train Communication Handover Models Based on DEA

Yuzhe Zhou
State Key Laboratory of Rail Traffic Control and Safety
Beijing Jiaotong University
Beijing, 100044 P.R. China
12120198@bjtu.edu.cn

Bo Ai
State Key Laboratory of Rail Traffic Control and Safety
Beijing Jiaotong University
Beijing, 100044 P.R. China
boai@bjtu.edu.cn

*Abstract*—Broadband communications for high speed train is becoming a main trend in high mobility communications. The main bottleneck of this communication network is handover, since the handover occurs so frequently and the delays are so long that broadband real-time communication cannot apply. Various handover models have been developed and studied recently. However, no comprehensive evaluation method for these models is employed. To this end, we borrow Data Envelopment Analysis (DEA) method to evaluate six typical handover system models. Handover models that to be evaluated are introduced. A brief presentation of DEA and its characters is provided. A specific procedure of the evaluation is proposed. Then the results of the evaluation are obtained by running the DEA. Finally, we give our comments and conclusions to all the handover models. We hope our work will supply a gap in the system evaluation area.

I.  INTRODUCTION

High speed railway is playing an important role in mass transportation all over the world. The development of high speed railways makes it more convenient for people's lives and works. Meanwhile, it puts forward various higher requirements on services of high mobility (i.e. above 300km/h) users. On the one hand, information of the train operation, the system state monitoring need real-time transmission to the control center, as a real-time dynamic information transmission feature of the railway network. On the other hand, with the development of social informatization, people need to stay connected to the network (especially the Internet) through a variety of communication terminals. Nowadays, wireless communication has a wide range application in high speed rails. Wireless access networks like satellite, GSM/GPRS, 802.11, and special wireless access networks are being used for train operation and passenger application. However, these access methods cannot meet the current or future requirements through the development of railways and communications, i.e. the high mobility, high data rate, real-time, and more reliability. Many new wireless access technologies such as WCDMA, WiMAX, and LTE, etc. have emerged. Each method has its own features with higher Quality of Service (QoS). However, they are not vastly applied in the railway communications. These new access networks are under research and field test for applying to high speed train (HST) communications. However, which wireless method is the best suitable one for the high speed rails? Evaluations of these methods should be provided.

Since the distribution nature of the high speed railway, the wireless access network for it also has the distribution feature, i.e. it covers the whole railway through a lot of cells. Therefore, the construction of distributive communication system is a huge project. The evaluation of this kind of system is quite different. Many factors may count for the evaluation, mainly the system performance and the investment. In the view of ensuring reliability and real-time performance of the system, it must ensure the underlying technology to be reliable and real time at first. In addition, these performances largely rest on the handover performance. So handover performance is the bottleneck of the wireless cellular networks in order to provide QoS for the users and to support users' high mobility. Evaluation of the system performance can be obtained by simply evaluate the handover model which involves only two adjacent cells. For the investment part, we can firstly divide the network according to its distribution feature and focus on the evaluation of one or two single cells, and then we put the evaluations together. Under this condition, we can consider these main factors jointly via evaluating the handover model. In the evaluation we consider the efficiency factor, the reliability factor, and the economic factor jointly. Note that trade-offs are often being made between efficiency and reliably factors. An evaluation method which can combine these factors is needed.

Data Envelopment Analysis (DEA) is an economic method to evaluate the cost/revenue of a system. In [1], DEA is used to evaluate the wireless communication sector's cost/revenue in different countries. And in [2], the authors employ DEA to make the choice of the best wireless access modes among GPRS, CDMA, and SCDMA which is going to apply in a certain city. The authors in [3] apply DEA to evaluate an Artificial Neural Network model. However, these evaluations mainly focus on the economic part and the results are references for management or construction. Indeed, they have given us a revelation that DEA can be used to evaluate various factors of various systems. In this paper, we chose DEA to jointly evaluate the combination of efficiency, reliability, and economic factors. And additionally, we combine objectivity and subjectivity in the evaluation, i.e. the quantitative and qualitative analyses are provided simultaneously. The purpose of this paper is to evaluate the performance and cost of various handover models for high speed train communications and give firm remarks of these access technics.

The rest of this paper is organized as follows: Section II introduces the DEA method. Section III gives a quick overview of handover models of high speed railway communication. Relative data of handover models are gathered and analyzed. Then a detailed DEA evaluation of these handover models is provided in Section IV. Finally, Section V concludes this paper.

## II. DATA ENVELOPMENT ANALYSIS

Data Envelopment Analysis is a data oriented approach for evaluating the performance of peer entities called Decision Making Units (DMUs) which convert multiple inputs into multiple outputs. We may want to know the overall performance of DMUs by their inputs consumed and outputs generated. It is desirable that we are able to combine various inputs and outputs into one measure. DEA provides this feature. Other alternatives such as analytic hierarchy process is too complicated and subjective. However, sensitivity to selection of inputs and outputs and number of variables is the main drawback of DEA. Through numerical results, we find that this sensitivity issue is negligible via peer comparison.

As introduced by Charnes, Cooper, and Rhodes, i.e. the CCR model [4], there are $n$ DMUs to be evaluated. Each DMU has $m$ different inputs and $s$ different outputs. Specifically, $DMU_j$ consumes amount $x_{ij}$ of input $i$ and produces amount $y_{rj}$ of output $r$. Assume that $x_{ij} \geq 0$ and $y_{rj} \geq 0$ and at least one positive input and one positive output. The ratio of outputs to inputs is used to measure the relative efficiency of the $DMU_j = DMU_o$ to be evaluated relative to the ratios of all of the $j = 1, 2, …, n$. We can interpret the CCR model as follows:

$$\max \quad \frac{u^T Y_0}{v^T X_0}$$
$$s.t \quad \frac{u^T Y_j}{v^T X_j} \leq 1, \quad j = 1, \cdots, n, \quad (1)$$
$$u \geq 0, v \geq 0.$$

where $u, v$ are $s \times 1$, $m \times 1$ weight factor vectors, respectively. $Y_0, X_0$ are output and input vectors of the evaluated DMU. The LP dual problems of the CCR model are referred as Farrell models and have the input oriented form or the output oriented form of:

$$\min \quad \theta$$
$$s.t \quad \sum_{j=1}^{n} X_j \lambda_j \leq \theta X_0, \quad (2)$$
$$\sum_{j=1}^{n} Y_j \lambda_j \geq Y_0,$$
$$\lambda_j \geq 0, j = 1, \cdots, n.$$

or

$$\max \quad \sigma$$
$$s.t \quad \sum_{j=1}^{n} X_j \lambda_j \leq X_0, \quad (3)$$
$$\sum_{j=1}^{n} Y_j \lambda_j \geq \sigma Y_0,$$
$$\lambda_j \geq 0, j = 1, \cdots, n.$$

The optimal values $\sigma$ can be less than, equal to, or greater than 1. Now we are capable to rank the DMUs according to their aggregated output to aggregated input ratios by $\sigma$. In Section IV, we apply input oriented and output oriented forms to evaluate.

The result of the evaluation is to explain the relative efficiency. A DMU is to be rated as fully (100%) efficient on the basis of available evidence if and only if the performances of other DMUs do not show that some of its inputs or outputs can be improved without worsening some of its other inputs or outputs. The efficiency of a system consists of three components: technical efficiency (TE), allocative efficiency (AE), and cost efficiency (CE). Technical efficiency reflects the ability of DMUs to obtain maximal output from a given set of inputs. Allocative efficiency reflects the ability of DMUs to use the inputs in optimal proportions. Cost efficiency reflects the ability of DMUs to produce a given amount of output with minimum cost.

We should point that the definition of a DMU is generic and flexible. To allow for applications to a wide variety of activities, we use the term DMU to refer to any entity that is to be evaluated in terms of its abilities to convert inputs into outputs. In the following part of this paper, we use DMU to refer to the handover model to be evaluated. According to the practical significance of the input and output variable of each DMU of the DEA method, that is an input value smaller and an output value bigger being better, which guarantees that the DMU has the highest benefit. Note that there is no limit to use various factors as the inputs and outputs. Then the performance metrics and costs of handover models can be applied to DEA.

## III. HST COMMUNICATION HANDOVER MODELS

Handover is the mechanism that transfers an ongoing call from one cell to another as a user moves through the coverage area of a cellular system. Handover model involves in structural network configuration and performance metrics. Besides, as one of the most important functionality of a mobile system, the handover procedure needs to be designed according to the nature of the network architecture. Therefore, different characterized access network architectures determine different handover models. In this section, six wireless access networks and related handover models for high speed train communication are reviewed and analyzed, and related data information is provided for evaluation in Section IV.

### A. Satellite to Train

Satellite communication has large coverage, and unlike the cellular network, much less ground equipment next to the tracks is needed. Almost no handover is needed due to the height of the satellite. However, blind spots exist in its coverage area where there are tall buildings, mountains, and tunnels. There is a method of applying the Wireless Local Area Networks (WLAN) as complement for loopholes of the satellite communication [5]. Yet its continuous coverage is much smaller than we had thought. So handovers are still required. In addition, satellite signal suffers high losses in bad weather conditions. Satellite communication links with limited bandwidth (typically 4MB) and high costs (satellite renting cost and user equipment cost) cannot meet the needs of a large number of passengers. What's more, satellite communication has a considerably round link delay (about 4 seconds for good), making it not suitable for real-time applications.

*B. LCX*

Leaky coaxial cable (LCX) can provide a uniform coverage of radio signals without mutual interference. It can reuse several frequency points in the long segment for HST with high bandwidth efficiency. The typical 100-400m-length LCX [6] is deployed close to the train, so the signal quality is good and transmission power is low. However, stringent requirements on the slot size, together with heavy loss of radio signal which should install a great number of repeaters to compensate for transmission losses, making the relay high cost. Its networking initial investment cost could be very high.

*C. RoF*

Radio over Fiber (RoF) [7] is a recently new framework which is very suitable for HST broadband wireless access and has a promising future. Multiple RAUs are located along the railway and connected to one control center via a fiber ring. The electrical signals at the control center are converted into optical signals and transmitted to RAUs via the fiber. Then they are converted back to electrical signals and radiated by the antennas from RAUs. The 60 GHz technology is applied to gain a very broadband (1GB) compared with current access technologies. The RoF model is cost effective: one control center can associate with many RAUs and these about-100m-radius microcells are equivalent, and the antennas are linear-radiated, so much transmission power is saved. Moreover, there is no need for handovers when the train moves from one RAU to another under the same control center. Thus, RoF model can reduce the number of handovers significantly. Besides, handover time can be reduced since optical switching time is typically $10^{-6}$s.

*D. RS-assisted*

Relay station (RS) is widely used in communication systems to increase capacity, expand network coverage, enhance weak-field zone, and even have an information gathering function. The purpose of the relay assisted handover model is to ensure signal coverage and strength to meet the minimum reliable communication requirements in the overlapping area. In this model [8], the RS is located in the middle of overlapping region and associated with the source Base Station (BS). The RS first act as a repeater and a diversity gain is achieved. When the handover triggering condition is satisfied, the RS starts its power control to maintain the connection of HST with source BS when the target BS is preparing handover. After the handover is completed, the RS stops its transmission.

*E. SFN*

In the Single Frequency Network (SFN), all cells of the cellular system operate at the same frequency. This is reasonable, since the whole train can be treated as a user terminal. Assuming that all the BSs in the SFN are synchronized, no new low-layer connection needs to establish. A good realization is to apply Coordinated Multipoint Transmission/Reception (CoMP) [9]. In this model, two adjacent cells form a Cooperative Transmission Set (CTS), where the two eNBs are cooperatively working. Same frequency is used to send the data to one train within the CTS, so soft handover is available at the cell boundary. CTS continuously reconstructs along with the train's moving direction. Only one frequency is used for one train, so the bandwidth can be much broader. Each train can achieve diversity gain and seamless handover.

*F. Dual-soft*

In the Dual-soft handover model [10], two antennas are installed in the front and the rear of the HST, respectively. When the HST is at the cell boundary, the front antenna performs handover to the target BS, and the rear antenna is still communicating with the source BS, so that the connection can be maintained during the entire handover procedure. Since bi-casting is the most basic data forwarding scheme to support fast handover in standard LTE system. The goal of this technique is to eliminate the data forwarding delay between the serving BS and the target BS. Dual-soft handover model takes advantage of bi-casting and dual antennas, then a soft handover can be effectively seamless, reducing the handover delay and execution overhead.

IV. HANDOVER MODES EVALUATION BASED ON DEA

The effectiveness of the various wireless access networks for high speed railway communication, a large part of that depends on the coverage of networks. In addition, the main problem of the network coverage is to move across different cells without disruption of communication, i.e. the handover issue. In this Section, we will focus on applying DEA to evaluate the effectiveness of the various handover models presented in Section III. And then we use the evaluation results to infer the effectiveness of relevant access networks. In our evaluation, each handover model is only being considered of communication between ground stations (or satellite) and HSTs, typically two adjacent cells, classic handover scenario. Considering the coverage of a base station, we can calculate the number of base stations which covers a specific length of railways. Furthermore, the effectiveness of the entire coverage can be calculated.

*A. Data Acquisition*

Table I shows the performance metrics and costs of the six handover models presented in Section III. All the data is acquired from the cited papers or related materials, and has been further processed (e.g. train speed is converted into handover rate). We take six factors for evaluation. Note that there is no number or type constraint to the choice of factors. For the readers' convenience, some items need to be explained. The handover rate item here refers to how many seconds between two handover triggers. It's calculated by the ratio of two middle points of adjacent overlapping areas to the train speed. The handover delay is the time from handover trigger to handover completion. The item cost is more flexible, since some of these models have not been realized yet, we just give our estimated values. The SNF and Dual-soft model only need slight change of adding special equipment to an HST. The RS-assisted model needs one relay station between two base stations. The RoF model needs to establish new RAU to form a microcell. These values can be modified if there are more accurate data, and the evaluation can be continued.

It is interesting to notice that the cost items in Table I are a little tricky, because they are not comparable. For example, the SFN and Dual-soft models only need new equipment on trains; the Satellite model has no base stations; other models have to establish new stations. And each coverage area of BS is not the same. So, we further average the cost values in Table I with respect to the coverage areas of each model, and then the averaged cost values in Table II is obtained.

TABLE II AVERAGE COSTS

| Modes\Items | Coverage per Cell | Cost/km |
|---|---|---|
| Satellite | 250km | 4 |
| LCX | 0.3km | 100 |
| RoF | 0.1km | 50 |
| RS-assisted | 4.8km | 2 |
| SFN | 4.8km | 0.2 |
| Dual-soft | 1.4km | 0.1 |

*B. DEA*

Through analyzing the distribution communication system, the evaluating model for wireless access system includes three kinds of factors: efficiency, reliability, and economic (green). The efficiency decides the capacity, data rate, and user-mobility adaptability of a communication system. This means the higher data rate, the more users it can hold, and the higher mobility the users are, the more efficient a system is. The reliability determines the success probability, seamless access and robustness of the communication links. The economic decides the construction investment and maintenance cost including the power consumption of the wireless access system. It is the main reference to evaluate the implementation of the model.

In the DEA, the six handover models are regarded as DMUs. Items such as cost, transmission power, and handover delay are inputs; Bandwidth, handover rate, and success probability are outputs. Because an input value smaller and an output value bigger are better. By setting up a linear programming model, a hypothetical synthetic model S is constructed based on the input and the output of the six handover models. A weighted average input (or output) of the six handover models is applied as the input (or output) of the hypothetical synthetic model S. In the constraints of this linear programming model, the output of the synthetic model S must be greater than or equal to the output of the model A (any one of the six handover models), and the input of S must be less than or equal to the input of A. If the input of S is less than the input of A, then the synthesis model has more output with less input. Therefore, model A is viewed to be relatively inefficient compared with the synthetic model S, which means A can be considered relatively inefficient than other handover models. Table III shows the results after running this DEA. We calculate both output oriented and input oriented form of CCR model. Three assumptions are made: First, only technical metrics are considered, no cost involved, this is a very ideal assumption. Second, cost values in Table I are considered with technical metrics. Third, averaged cost values in Table II are considered with technical metrics.

*C. Model Evaluation*

We first analyze the output oriented DEA results. The results are obtained via solving the relevant DEA models using (3) presented in Section II. The lower the factor is, the better the output relative efficiency of the model is. Note that things turn out to be different under different assumptions. For the technical only assumption, Satellite model has a score of 3, which means that the hypothetical synthetic model S can use the same input as the satellite model but has an output of 2 times bigger than the satellite model. So the satellite model is relatively inefficient. Same comments to RS-assisted model, SFN model and Dual-soft handover model. In addition, their technical performance is bad. The LCX model and the RoF model are output relatively efficient, which means they perform well in converting inputs to outputs, i.e. overall technical metrics are properly achieved. When taking the cost into account, the SFN and Dual-soft handover models become relatively efficient. The RS-assisted and satellite models are still relatively inefficient. Anyway, the RS-assisted model improves its score considering the cost. Since the investment of satellite model is so large that it is not capable of improving its score. When considering the average costs, all the six models are efficient. This means that in order to apply new models to substitute the existing one, the average cost in the long run must be considered. Once the scores are all the same, subjective evaluation can be employed. One or several metrics that we are especially interested in can be chosen to make the final decision. For example, because we are considering the handover performance on high mobility scenario, the handover delay overruns other performance metrics. Comparing the handover delay column in Table I, we see that RoF model performs the best, so RoF is chosen to be the most appropriate handover model for the high speed train communications.

Now, let's see the input oriented DEA results. The results are obtained via solving the relevant DEA models using (1) presented in Section II. The higher the score (no more than 1) is, the better the input relative efficiency of the model is. There are three efficiency measures which are presented in Section II. The RoF model is still the best model. For the allocative efficiency in technical only assumption, SFN model has a better score than the other four models, which means it has its inputs in better proportions. And the satellite model has its inputs in the worst proportions. Note that CE=TE AE. When it comes to the cost, the investments of satellite, LCX, and RS-assisted models are not appropriate for their performance, i.e. investments are higher, and performance improvements are smaller. SFN and Dual-soft models take better advantage of the investments. For the average cost assumption, conclusions may be different from the output oriented results. LCX and RS-assisted models still have bad AEs, indicating they are not allocative efficient, while satellite, SFN, and Dual-soft models are doing well.

After the score analysis, we give our comments on the six models. Satellite has a bad technical efficiency but a good averaged cost efficiency, which means that it does not meet the requirements of the broadband high speed train communications, but in the long run, its cost does not seem unaffordable. The existence of Thalys in Europe may have proved that satellite model is still alive. LCX model has a higher technical efficiency but a lower averaged cost efficiency, which indicates that the LCX network coverage solution is not suitable for the whole line of HST. Instead, it is an alternative to other wireless access methods in tunnels or other specific conditions. Japan has applied this model to the Shinkansen. Besides the existing models, the other four models are still under research or test. RoF performs very well in our evaluations. Before we argue that it is the most appropriate handover model for high speed train communication, more accurate cost assessment must be further made. RS-assisted model can improve some technical performance metrics, however, the investment of establishing many new relay nodes will cost too much. It is not appropriate for the coverage of the whole railway line. Instead, it is a good solution in weak fields along the railway. SFN and Dual-soft do not improve much of the model performance. Anyway their investments are relatively low compared with the RS-assisted model. So before they are chosen to be applied in the high speed rail communications, improvements and further evaluations should be made.

Note that, we can also apply a dynamic DEA model to evaluate the model efficiency. Since the technologies are developing, handover model performance metrics are changing. Dynamic DEA will give a time varying evaluation, and the improvement will be made only if a metric is beyond a certain range of the score in the static DEA evaluation. This kind of model needs more specific data, and is more complicated. We hope our work will give some valuable information to our colleagues.

## V. CONCLUSION

In this paper, we apply data envelopment analysis to evaluate the existing and novel handover models of high speed train communications. The difference of our work and the existing system evaluation methods is that we combine all the system metrics including efficiency, reliability, and economic metrics into one evaluation model. The former works mainly considered the economic metrics such as investment and maintenance cost to decide whether a system should be established or should be substituted by another new system. However, our proposed work expands this method to evaluate various technical metrics independently, or together with the economic metrics. We mainly use this method to evaluate the communication system, especially the handover models under the high speed railway communication trend. The evaluation results show that the RoF model is the best appropriate system to support high mobility communications. SFN and Dual-soft models also perform well. LCX and RS-assisted models can be employed in special situations. Satellite can still remain existence since direct links can be made to the train, so it would be a standby system in case of some ground systems were destroyed. Note that the DEA method can also apply to evaluate other kinds of systems or subsystems. It can also unify various metrics what we are interested in into one evaluation model. However, since a good evaluation needs much more specific and accurate data, our models do have limitations in view of the data acquisition. Anyway, we hope the methods and ideas use in our work can be enlightening and supply a gap in system evaluations.

ACKNOWLEDGMENT

The authors would like to express their great thanks to the support from the 863 Plan of China under Grant 2011AA010104, Beijing Municipal Natural Science Foundation under Grant 4112048, the project of State Key Lab under Grant RCS2012ZT013, the Key Project of Chinese Ministry of Education under Grant 313006, the Fundamental Research Funds for the Central Universities under Grant 2010JBZ008 and the project of State Key Lab under Grant No. RCS2011ZZ002.

TABLE I MODEL DATA TO BE EVALUATED

| Items Modes | Cost (10000 RMB) | Channel Bandwidth (MB) | Transmission Power (W) | Handover Rate (s) | Handover Delay (s) | Success Probability |
|---|---|---|---|---|---|---|
| Satellite | 1000 | 4 | 30 | 3000 | 4 | 0.95 |
| LCX | 30 | 2 | 0.5 | 2.5 | 0.1 | 0.95 |
| RoF | 6 | 1000 | 1 | 300 | 0.005 | 1 |
| RS-assisted | 10 | 1 | 42 | 30 | 0.1 | 0.95 |
| SFN | 1 | 10 | 40 | 40 | 0.5 | 0.97 |
| Dual-soft | 1 | 4 | 80 | 15 | 0.4 | 1 |

TABLE III DEA RESULTS

| Efficient Score Models | Output oriented | | | Input oriented | | | | | | | | |
|---|---|---|---|---|---|---|---|---|---|---|---|---|
| | *Technical Only* | *Cost* | *Average Cost* | *Technical Only* | | | *Cost* | | | *Average Cost* | | |
| | | | | TE | AE | CE | TE | AE | CE | TE | AE | CE |
| Satellite | 3 | 3 | 1 | 0.333 | 0.084 | 0.028 | 0.333 | 0.152 | 0.051 | 1 | 1 | 1 |
| LCX | 1 | 1 | 1 | 1 | 0.396 | 0.396 | 1 | 0.194 | 0.194 | 1 | 0.104 | 0.104 |
| RoF | 1 | 1 | 1 | 1 | 1 | 1 | 1 | 1 | 1 | 1 | 1 | 1 |
| RS-assisted | 21.1 | 1.96 | 1 | 0.095 | 0.556 | 0.053 | 0.516 | 0.27 | 0.139 | 1 | 0.268 | 0.2681 |
| SFN | 42.7 | 1 | 1 | 0.024 | 0.911 | 0.022 | 1 | 1 | 1 | 1 | 1 | 1 |
| Dual-soft | 80 | 1 | 1 | 0.025 | 0.579 | 0.014 | 1 | 0.904 | 0.904 | 1 | 0.885 | 0.885 |